# Artificial Atoms Based on Correlated Materials

J. Mannhart, H. Boschker, T. Kopp, R. Valentí

Max Planck Institute for Solid State Research, 70569 Stuttgart, Germany

Center for Electronic Correlations and Magnetism, Experimental Physics VI, Institute of Physics, University of Augsburg, 86135 Augsburg, Germany

Institute of Theoretical Physics, Goethe University, 60438 Frankfurt am Main, Germany

**Abstract:**

Low-dimensional electron systems fabricated from quantum matter have in recent years become available and are being explored with great intensity. This article gives an overview of the fundamental properties of such systems and summarizes the state of the field. We furthermore present and consider the concept of artificial atoms fabricated from quantum materials, anticipating remarkable scientific advances and possibly important applications of this new field of research. The surprising properties of these artificial atoms and of molecules or even of solids assembled from them are presented and discussed.

*Pronouncements of experts to the effect that something cannot be done always irritated me.*

*Leo Szilard, 1963*

**1. Correlated Electron Systems**

Based on the outstanding progress of materials technologies, in particular of thin-film technologies and of wet-chemical chemistry, a wide range of low-dimensional electron systems has become available for exciting scientific research and many remarkable applications. The two-dimensional (2D) electron gases (2DEGs) in semiconductor heterostructures are systems of special importance in this context. These 2DEGs, used for example in the high electron mobility transistors of cell phones and in a variety of semiconductor lasers [1], are the basis of the integer and the fractional quantum Hall effects [2-4]. Also zero-dimensional electron systems, called quantum dots, have become widely used, for example as markers for imaging biological systems such as cells or even organisms, or as color pixels in displays [5]. Essentially all of these electron systems are derived from uncorrelated, mean-field type materials such as Si or III-V heterostructures. The reason for using "mean-field compounds" is primarily historical. With the rise of the semiconductor technology, for example, it was already possible in the 1970s to grow semiconductor heterostructures [6]. Since then, a huge variety of low-dimensional electron systems has been fabricated with great success from semiconductors, and electron mobilities as high as $36 \times 10^6$ Vs/cm$^2$ have been achieved [7]. Even in such archetypical semiconductors can the behavior of low-dimensional electron systems be altered by correlation effects, which occur if the energy of the effective interaction energy $E_{int}$ between two electrons is at least comparable to the kinetic energy $E_{kin}$. For many electron systems, the direct Coulomb interaction $U(r)$ provides the main component of $E_{int}$, to which, however, exchange energies $J$ or indirect interactions via exchange of bosonic excitations or of multiple particle-hole excitations may also contribute. The kinetic energy $E_{kin}$ can frequently be characterized by the electrons' bandwidths $W$.

We identify four ways or mechanisms according to which electron systems are correlated or by which correlations may be generated. These mechanisms are illustrated in Fig. 1.

First, we note that correlations are already induced in electron gases if their electron density $n$ is sufficiently reduced. In the simplest case, the ratio $E_{int}/E_{kin}$ of 2D electron systems is proportional to the Wigner-Seitz radius $r_s = (\pi n a'_B{}^2)^{-1/2}$, where $a'_B$ is the effective Bohr radius $a'_B = \varepsilon_{eff} a_B / m^*$, with $\varepsilon_{eff}$ being the effective dielectric constant of the electron system, $a_B$ the Bohr radius, and $m^*$ the effective electron mass measured in units of the free-electron mass. For 3D systems the Wigner-Seitz radius equals $r_s = (4\pi n a_B{}^3/3)^{-1/3}$. When electron gases are diluted to smaller $n$, for example when depleted by a gate field, they become correlated at sufficiently small *n* (for reviews see [8,9]) (Fig. 1b).

Second, the kinetic energy of 2D systems is quenched by magnetic fields applied perpendicular to the 2D plane that are large enough to force the electrons into Landau orbits. Typically, field strengths of several Tesla are required. The effective bandwidth $W_L$ of the individual Landau levels is extraordinarily small, say < 100 µeV, so that $W_L \ll U$ and electronic correlations are relevant. This approach, illustrated in Fig. 1c, is used to induce the electronic correlations associated with the fractional quantum Hall effects [10]. It is applicable only to 2D systems, because the electron motion parallel to the magnetic field has to be suppressed to protect the gaps between the Landau levels. The third mechanism to induce correlations is structuring the electron system to induce quantum-wells in one, two, or three dimensions, using for example superlattice structures or lateral patterns to generate minibands in the valence or conduction bands, as is well-known from semiconductor devices (see, *e.g.*, [11,12] (Fig.1d). The bandwidths of the minibands and therefore the kinetic energy of the electrons is small, which fosters correlations. If induced in such a manner, the correlation parameters can be tuned by varying the design of the pattern and superlattice stacking. In the fourth general mechanism by which electron systems are correlated, large ratios of $E_{int} / E_{kin}$ are intrinsically provided by the material itself, which may for example be characterized by large local Coulomb repulsion energies *U* on ionic sites and small bandwidths *W*. This situation, depicted in Fig.1e, is the one on which we will focus in the remainder of this article.

Allowing for a broad material base, we here consider among other compounds low-dimensional electron systems fabricated from materials that are already inherently correlated, *i.e.,* materials for which $E_{int} \gtrsim E_{kin}$, (see Fig. 1e). There exists a rich variety of correlated compounds [13,14], such as oxides, nitrides and organic materials to name just a few families, many of which also feature a wide spectrum of functional properties [15,16]. Owing to enormous progress in thin-film technology, particularly in molecular beam epitaxy and pulsed laser deposition controlled by reflective high-energy electron diffraction (RHEED), 2D systems based on correlated materials have been realized and are beginning to be investigated. For many of these compounds, some properties differ fundamentally from those of canonical semiconductors [17,18]. In the following we describe several key examples of such 2D systems:

At the interface between $LaAlO_3$ and $TiO_2$-terminated $SrTiO_3$, a well-conducting 2D electron system is formed [19] if the $LaAlO_3$ layer is at least 3 unit cell layers thick [20] (see Fig. 2). In this system the mobile charge carriers reside in Ti 3*d* orbitals. Measurements of the interface spectral density of states by scanning tunnel spectroscopy [21] indeed revealed the splitting of Ti-3*d* Hubbard bands at the interface, in agreement with LDA+*U* calculations based on a *U* value of 2 eV at the Ti site. This electron system is a 2D electron liquid rather than a 2DEG [21]. Jang *et al.* analyzed the conductivity of heterostructures consisting of unit-thick layers of several rare-earth oxides (La, Pr, Nd, Sm, Y) embedded in epitaxial $SrTiO_3$. Their results indicate that the interfacial conductivity is dependent on electronic correlations that decay spatially into the $SrTiO_3$ matrix [22]. Comparable correlations have been found for the electron system in $SrTiO_3$ quantum wells embedded in $GdTiO_3$-$SrTiO_3$-$GdTiO_3$ heterostructures. In these studies, the electron liquid behavior has been derived by analyzing the temperature dependence of the quantum well's resistivity, which shows a $\rho(T) = \rho_0 + AT^2$ dependence, where $A$ is proportional to the correlation-induced mass enhancement of the charge carriers [23]. Also, the 2D electronic system induced on the bare $SrTiO_3$ (001) surface has been found by angle-resolved photoemission spectroscopy to be an electronic liquid, with correlations being evidenced by spectral weight well above the bottom of the conduction band at the $SrTiO_3$ surface. This spectral weight is associated with a finite electron self-energy at high binding energies, despite a small effective mass of only 0.5-0.6 $m_e$ [24]. Electronic correlations were explored by transport and optical conductivity measurements [25,26] in very thin $LaNiO_3$ films as well, which were found to be low-dimensional Mott materials. Finally we note that unit-cell thick films of $SrRuO_3$ embedded in $SrTiO_3$ have been found to be magnetic and conducting – yet another example of a two-dimensional correlated system [27].

Two-dimensional and one-dimensional electron systems fabricated from correlated matter generate many relevant, exciting questions. For example, which effects will be shown by 1D conductors fabricated from correlated materials? What spectrum of properties and phenomena will the integer and fractional quantum Hall effects display if they are realized in materials with strong electronic correlations? In this manuscript, however, we will focus on zero-dimensional electron systems and ask which properties we may expect from artificial atoms fabricated from correlated materials.

## 2. Conventional Quantum Dots and Artificial Atoms

Zero-dimensional electron systems, aka quantum dots, were introduced in 1981 by A. Ekimov [28] who investigated CuCl and CdSe nanoparticles. Pioneering contributions were also made by A.L. Efros [29], L.E. Brus [30], A. Arakawa, and H. Sakai [31]. Size quantization of the electronic states affects the electron and exciton energies in these dots. Size quantization occurs if the characteristic size of a dot does not exceed by much the Fermi wavelength of the electrons or the Bohr radius of the excitons. These size quantization energies increase with shrinking dot size. The resulting electron states typically form a shell structure with an energy spectrum that mimics those of natural atoms. The electron system of the dot is described by a coherent many-body wave function with one macroscopic phase. Because of the shell structure of the electron states and their coherency across the dots, quantum dots are appropriately described as "artificial atoms" [32-35].

Here, we use the quantum phase coherence as the defining criterion for "artificial atoms", because (i) the quantum coherence entails an orbital structure of the electron states, and (ii) phase coherence is required to be the basis of characteristic properties of atoms. For example the electronic structure must be adequately presented by a solution of the Schrödinger equation [36]. In some of these dots the phase coherence length seems to exceed the dot size at room temperature, so that even under ambient conditions the electronic states of the dots are coherent with an orbital character described by the Schrödinger equation. Such artificial atoms have been fabricated from a large variety of compounds, including metals, ionic compounds, and even nanotubes (see, *e.g.,* [37,38]). Artificial atoms have also been realized using 2DEGs in semiconductor heterostructures, typically by using gate structures (for overviews, see, *e.g.*, [39,40]) as well as superconducting circuits (for an overview see [41]). An exciting feature of these "designer" atoms is that their electronic properties are tunable, which has been exploited in many experiments and applications. Their properties are susceptible to the geometry of the dots, to pressure, and, importantly, to electric fields applied by gate electrodes. Gate fields are applied to study and exploit Coulomb-blockade effects that arise from single-electron charging of the dots and are used, for example, for electron counting. The materials of these quantum dots are mean-field electron systems. Electronic correlations, however, have been induced into such dots and studied in detail by adding magnetic scatterers acting as Kondo impurities [42] and by quenching the kinetic energy by inducing with magnetic fields Landau orbits [43]. Quantum dots have already been finding their way into applications such as markers for imaging biological systems or as color pixels for displays [5]. For their phase-coherent properties, some of these quantum dots are also of interest for applications in quantum cryptography and, hypothetically, in quantum computing. Note that confinement effects are also being pursued as a means to enhance the critical temperature of superconducting quantum dots [44,45]. Excellent overviews on quantum dots are, for example, provided by [33,35,46,47,39,40].

## 3. Artificial Atoms from Correlated Materials

In the following we consider artificial atoms, molecules, and solids made from correlated materials, which as described above exist in large varieties and with many different properties. We also examine artificial atoms made from other types of so-called quantum matter, such as compounds with useful functional properties that are not necessarily derived from electronic correlations (see Fig. 3 for an artist's rendition of such a dot). Which properties could such atoms and their assemblies possibly have? Which properties can be achieved with artificial solids built from such atoms? Which phenomena could be expected if it were possible to fabricate phase-coherent nanostructures from such artificial atoms? The possibilities are tremendous, and a broad spectrum of correlated materials with remarkable properties exist from which such artificial atoms can be built, including chalcogenides, heavy fermion systems and organic materials, to name but a few examples. Many of these materials can be prepared as nanoparticles (see, *e.g.*, [48,49]) or with atomic-layer precision as heterostructures [50]. These heterostructures may be grown in a self-organized manner to form nanostructures (see, *e.g.,* [51]), or be patterned to length scales of tens of nanometers and below. Although many of the proposed structures have yet to be achieved, the current progress of nanoparticle synthesis (see, *e.g.*, [48,49]) and deposition and patterning technologies of quantum matter (see, *e.g.*, [52]) suggests that a broad range of such artificial atoms will be realized in the next few years. For virtually all of these systems, the phase-coherence length has not yet been measured. It is therefore not known out of which the creation of phase-coherent quantum dots of practical sizes will be feasible, let alone the fabrication of phase-coherent assemblies of several quantum dots. Yet, it seems realistic that at least some of these materials can be grown with defect densities low enough to yield low-temperature inelastic mean free paths of tens of nanometer (for single particle transport). Indeed it has already been demonstrated that correlated electron systems can be structured to this length scale. In the 2D electron liquid at the $LaAlO_3$-$SrTiO_3$ interface, for example, the observation of weak localization and universal conduction fluctuations was reported for devices with lengths as large as 4 μm [53-55], while 5-nm-wide lines and dots with diameters as small as 2 nm have been written with a charged AFM tip [56]. Comparably small, gated quantum dots in $LaAlO_3$-$SrTiO_3$ [57] were recently used to analyze the carrier charge in the normal state of $SrTiO_3$, the charging behavior of the dots suggesting a value of twice the electron charge [58]. Also electron beam lithography has been successfully used to nanopattern correlated systems, with smallest features of ~50 nm currently produced in the $LaAlO_3$-$SrTiO_3$ electron liquid as illustrated in Fig. 4 [59-62]. Is it realistic to expect that quantum-matter-based dots can be fabricated in adequately small sizes to benefit from the phase coherence of many-body states? Indeed, for fundamental reasons the characteristic energy and length scales of these dots differ greatly from the length scales that characterize canonical semiconductors. As the effective masses and dielectric constants of these complex materials are characteristically large, the size quantization energies and charging energies are correspondingly small. For a given dot size, these energies are an order of magnitude smaller than those of canonical semiconductors. We therefore find for these materials a large window of temperature and dot size, for which typical correlation energies are the dominant energy scale and for which coherent electron systems can be expected next to the regimes in the dot-size temperature parameter-space in which the size quantization and charging energies dominate (see Fig. 5). Such artificial atoms will allow the study of coherent, correlated electron systems in experiments that are neither dominated by Coulomb blockade nor by finite-size effects. As we will show, these phase-coherent systems can be expected to reveal remarkable properties.

In the following we list key properties of such artificial atoms and explore important questions and problems. We explicitly note that some of these properties are specific to patterned heterostructures, because their atomic-layer sequence can be readily designed and controlled. Most concepts, however, also apply to phase-coherent nanoparticles and their clusters grown by chemical routes, say, from colloids (see, *e.g.*, [48,49]).

(a) We first neglect size quantization and consider that, due to the small size of the artificial atoms, the boundaries of the dots are relevant in controlling the characteristics of the materials from which they are fabricated. Translational and rotational symmetries are broken by the quantum dot's surface, the crystal lattice changes, also by relaxation, crystal fields are modified, bandwidths and electronic dispersions are usually reduced, and electronic and magnetic susceptibilities as well as Madelung energies are altered. Indeed, even lattice instabilities may develop or be suppressed. As a result, the fundamental correlation parameters of the material are modified, including altered and spatially-dependent Coulomb energies $U(r)$, charge-transfer energies $\Delta(r)$, hopping parameters $t(r)$, and exchange constants $J(r)$. These effects emerge in addition to changes of charge redistributions, built-in potentials, electric and magnetic fields, susceptibilities, defect populations, instabilities and phase separations, and are reminiscent of the effects that place at macroscopic interfaces involving correlated electron systems [63,64, 17]. Furthermore, also the kinetic energy of the electrons is a function of the dot size. The kinetic energy is small for tiny dots, favoring correlations. All these effects occur on their own length scales of typically a few unit cells. Possibly strong spatial dependencies of the electronic parameters lead to large gradients, also in high order, and add to the nonlinear, coupled behavior of the basic parameters.

It is noted that because the electron systems are confined into small particles, size-quantization effects are generated (see, *e.g.*, [65]). In a nonlinear manner, size-quantization may weaken or as well strengthen the modifications of the electron system just described.

As a result, the electron systems of quantum dots made from correlated materials may have characteristic wave functions, phases, ground states, excited states, catalytic functions, diffusion characteristics, and non-equilibrium properties that are unavailable from any other mean-field or bulk correlated system and may genuinely differ from those.

Owing to the altered materials' properties described here and the quantum confinement, spectra of electron states are induced in these quantum dots, which are tunable by the shape of the dot, epitaxial strain, and gate fields, as is the case for the quantum dots fabricated from mean-field systems. Together with the heterostructure materials and their stacking sequences these parameters can be varied to "design" the wave functions of the quantum dots. The spectra of energy levels influenced by the correlations will be apparent in the optical excitations and transitions.

Here it is important to recall that the characteristic properties and functionalities of the materials from which the dots have been fabricated are retained in the limit of large dots. Therefore, the quantum dots will be characterized in many cases by altered, distorted, weakened or even strengthened functional properties of these materials, such as magnetism, ferroelectricity, and multiferroic behavior. Explorations of the functional character as a function of increasing quantum dot size will illuminate the emergence of these many-body properties, starting from the unit-cell level.

(b) Artificial atoms of correlated materials are also potential building blocks of superconductors with optimized properties. Tuning the electron systems by using the effects described in (a) opens the door to altering and optimizing superconducting behavior by using small particles (see, *e.g.* [66-68,44,69]). It is also a compelling approach to use interfaces to enhance or induce superconductivity [70-83]. Here we mention two recent examples: (i) the coupling of a metallic or superconducting layer to a dielectric to optimize independently the host materials of superconducting charge carriers and their glue [76] and (ii) the coupling of an insulating layer with a strong local pairing to a (normal) metallic layer [77]. In a regime with intermediate kinetic transverse coupling (interlayer hopping), the strong phase fluctuations are sufficiently reduced and can enhance $T_c$. This approach has been explored in cuprate bilayers with a $T_c$ well beyond the corresponding bulk $T_c$ [79]. It is conceivable that such ideas can also be realized using arrays of artificial atoms, for example of high-$T_c$ cuprate composition, which are embedded in a host environment such that the interfaces are created and superconducting coupling is achieved among them (Fig. 6).

(c) The electron system of artificial atoms is characterized by a coherent many-electron wave function, because by definition, phase-breaking scattering processes need to be negligible to obtain a sufficiently large phase coherence length. For complex electron systems such as for example spin fluids it is interesting to explore what types of coherent states develop in artificial atoms and possibly in phase-coherent arrangements of many atoms.

If an artificial atom is coupled to an incoherent environment, provided for example by the substrate or electric contacts, a finite phase stiffness is also required to maintain the phase coherency. The phase stiffness necessitates a finite condensation energy of the multi-electron wave function and therefore requires suitable electron-electron interactions. These interactions may be indirect, using for example exchange bosons such as phonons or spin excitations, or direct, and they will enhance the phase coherence length. For quantum dots based on correlated materials, the use of electronic correlations to enhance phase stiffness may be possible and advantageous. Such a state with phase-stiffness will display the Meissner effect and, therefore, it will also be superconducting, in addition to possibly showing unusual many-body properties.We note that the requirements for phase stiffness could possibly be relaxed when experiments are performed on a time scale shorter than the decoherence time of the states.

Owing to phase coherence, the electrons in the dots are entangled and the dots may be characterized as qubits. Interestingly, the correlations in these qubits will allow us to realize excitation spectra that are not attainable otherwise. Here the question arises whether the correlations, which in many cases will enhance decoherence of the qubits, can be utilized in appropriately tailored structures to reduce decoherence. Another interesting issue concerns the behavior and application potential of qubits in which the individual states are coupled to functional properties.

(d) Following the analogy with natural atoms and conventional quantum dots, we now consider the possibility to couple two or even more dots into larger, coherent assemblies characterized by many-body electron wave functions with a quantum phase. Here we note that fabricating such molecules from semiconductor quantum dots is a current field of research and has encountered several challenges. It is nevertheless rewarding to consider which properties these assemblies would display if it were possible to fabricate them from correlated systems, even though their practical realization may not be possible with the means available in the near future. In such assemblies, the quantum dots and their links can—at least in principal—be individually gated such that the dots, the links, and the assemblies can be controlled, fine-tuned, or switched on and off. As interfaces are prone to inelastic scattering, such assemblies are described by several phase-coherence lengths: the phase-coherence lengths of the individual atoms, the effective phase-coherence lengths of the links, and that of the whole assembly. To achieve large values for the latter, which in many cases will probably prove to be challenging, it is therefore essential to couple the dots with as clean interfaces as possible, i.e. with as little inelastic scattering as possible. If different materials are combined, those combinations with matching electron states at the chemical potential will more easily hybridize and couple. In the case that the dots are strongly coupled and their wave functions are well hybridized, new molecules or solids are obtained, which, being phase-coherent, are characterized by one macroscopic superconducting wave function. Such metastructures, which pose intriguing multi-scale problems, are especially interesting if the molecules are built of different kinds of artificial atoms that possibly have different order parameters of complex ground states or different functionalities. Obviously, the freedom to design the geometry of arrangements of artificial atoms is virtually unrestricted, so that photonic lattices, metamaterials, and frustrated lattices can also be envisioned.

(e) A lattice of phase-coherent, correlated quantum dots (Fig. 7) is a solid-state analog of cold-atom systems, where complex artificial atoms form a phase-coherent ground state. In most cases this ground state, if achieved, is a fermionic condensate. Nevertheless, the electron systems in the dots may also be paired in the relevant energy range, giving rise to bosonic condensates. Low-dimensional electrons systems are not necessarily fermionic or bosonic, but may have an anyonic character, in which case anyon condensates may be formed [84]. In addition, it would seem that condensation of quasiparticles, e.g., of excitons, is possible in such lattices. Succinct differences exist between standard cold-atom systems and cold-atom systems built from artificial atoms. They differ first of all in terms of the size, density, character, variety of atoms, freedom of design, and ease with which disparate structures that, for example, comprise altered atoms or designs can be prepared. Condensates of conventional cold atoms are characterized by a superbly regular lattice and by the absence of defects. Defects, however, will always be inherent in condensates of artificial atoms; they are simply unavoidable due to growth and patterning irregularities. Substrates will typically be used for the growth and support of artificial-atom arrays, and will usually degrade their properties by adding noise and decoherence. It is an intriguing aspect, however, that on a larger length scale well-characterized defects can be easily designed and inserted as active centers into arrays of artificial atoms of correlated matter. It is in principle possible to prepare irregular lattices or to introduce disorder of different character and strength, such as by preparing systems with missing dots, anti-site defects, or selected dots composed of altered materials (compare, e.g., [85]).

(f) Recently it was shown that random, interconnected, nanoparticle networks can have a rich emergent behavior. In that case, each nanoparticle acts as a single-electron transistor with highly nonlinear characteristics, and the ensemble can be configured into a Boolean logic circuit by tuning static control voltages [86]. To achieve this, some hundred nanoparticles are needed to fabricate a universal logic gate. Such a number is necessary because the system requires a minimum degree of complexity. Artificial atoms based on correlated materials, however, are inherently complex. We therefore expect networks of such functional dots fabricated from quantum matter to show a richer emergent behavior, which thereby yields logic gates with smaller networks and more energy-efficient computation. In addition we anticipate possible data storage applications for complex networks as well as for single dots, which offer themselves for data archiving based on chemical modifications or phase transitions induced in the quantum matter.

The group of J. Levy presented the suggestion to use arrays of $LaAlO_3$–$SrTiO_3$ quantum dots written with conducting atomic-force microscopy tips as quantum simulators, for example to solve the Hubbard model [57,87]. Whereas correlation effects in $LaAlO_3$–$SrTiO_3$ are not needed for this simulation and have not been considered in this study, the suggestion provokes the intriguing question whether the correlation effects of quantum-dot arrays made of correlated materials would allow one to achieve condensation into correlated wave functions that are unattainable with mean-field, material-based artificial atoms.

(g) Further extending this vision of correlated quantum dots, we point out that, by altering the electronic states in the dots, for example by using staggered gate electrodes driven by appropriate voltage sequences, the electronic states of the dots may effectively be moved. This concept is illustrated in Fig. 8. For example, staggered gates permit moving one electron system around another one, swapping the positions of two electron systems, or moving a dot away from another dot *A* to make contact with a dot *B*. Such capabilities are valuable, for example, for exploring anyonic electronic systems or systems comprising Majorana-type quasiparticles, in particular because the resulting phase changes are accessible to measurements by using associated tunnel junctions.

(h) Lithographically it is straightforward to fabricate ever larger arrangements of such artificial atoms in one, two, and three dimensions. Furthermore, the large spectrum of available materials yields the possibility to couple functional dots into selected bosonic modes. Experiments analyzing the quantum mechanical coherence and the onset of incoherent, classical behavior of such systems as a function of size may therefore provide a viable road to study the transition from phase-coherent quantum physics to the classical world.

Obvious and key difficulties in fabricating such dots are posed by the need to pattern at the required length scales. For typical materials, size quantization and Coulomb blockade effects require dots smaller than 20 nm, whereas phase coherence may already well be achieved with larger dots (see Fig. 5). These length scales may be reached by electron-lithography-based processes [88,54,59,89,61,62] and, as already well demonstrated, by scanning probe writing [56]. The phase-coherence length of non-superconducting systems is controlled by the inelastic scattering length, inelastic scattering being caused, for example, by phonons and defects, and may never achieve the large values attainable in mean-field, canonical semiconductors or metals. It is therefore as crucial to grow materials with a low density of inelastic scatterers as it is to perform high-resolution patterning. It will be even more difficult to couple artificial atoms coherently via links, and it will be interesting to see which clever designs will be developed to achieve such linking.

## 4. Summary

We have sketched the vision of artificial atoms fabricated from complex and possibly correlated materials, generally known as quantum materials. These atoms are the ultimate nano-building blocks of quantum matter, for they contain the key physical elements of the complexity, emergent properties, and functionality of the corresponding bulk materials. For example, they might be fabricated as nanoparticles by chemical means, or with even more freedom of design by film deposition and patterning as heterostructures with layer sequences designed on an atomic-layer level. While the atoms themselves are remarkable systems to study coherent correlated electron systems, the emergence of functions and the role of defects, such atoms coupled into artificial molecules, phase-coherent arrays and further nanostructures offer ample opportunities for research and applications—if they can be achieved at all. Which properties and uses of these artificial atoms and their metastructures can we anticipate? Our attempts to answer these questions are limited, not only due to the space limitations of this article and our lack of imagination, but also because relevant effects and applications will probably only be discernible once these artificial atoms have been adequately explored and implemented [90].

## Acknowledgements

We gratefully acknowledge illuminating and inspiring discussions with many colleagues, in particular with Ali Alavi, Antoine Georges, Peter Horsch, Wolfgang Ketterle, Klaus v. Klitzing, William Phillips, Doug Scalapino, Jürgen Weis, and Peter Wölfle. We thank Carsten Woltmann for the photos shown in Fig. 4 as well as Jonathan Mannhart for support with Figs. 3 & 7. This work was supported by the DFG through TRR 80 and a Leibniz-grant.

**Figures**

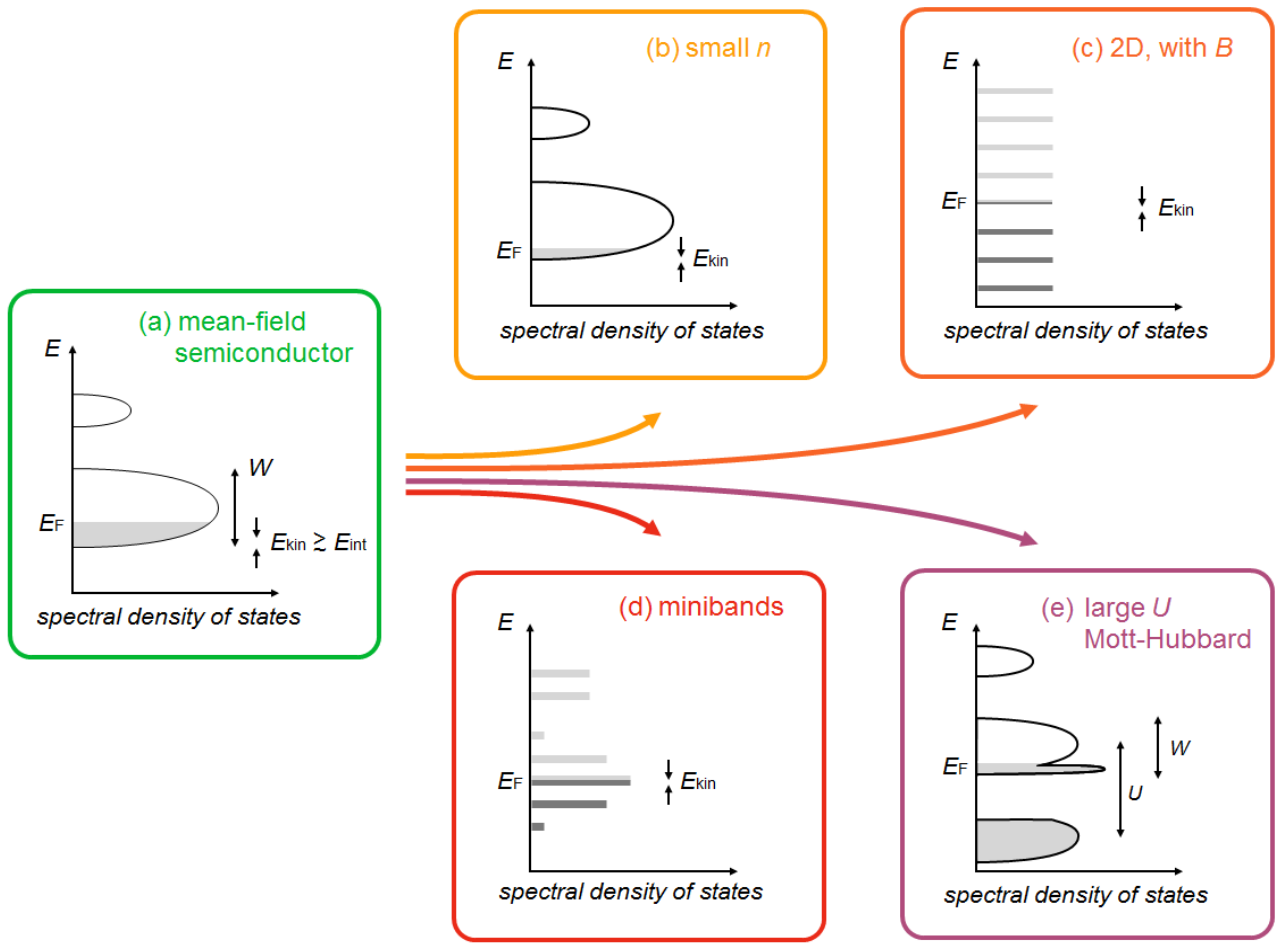

Figure 1

Fourfold way to electronic correlations. Starting from a mean-field semiconductor (a) characterized by a kinetic energy $E_{kin}$ being larger than the electron-electron interaction energy $E_{int}$, by a bandwidth $W$, and a carrier density n, correlations are induced by reducing $n$ to a level that $E_{kin}$ and $E_{int}$ become comparable (b). A reduction of $E_{kin}$ to values comparable to $E_{int}$ is also caused in 2D-electron systems by applying a magnetic field perpendicular to the 2D plane, as such a field collapses the original semiconductor bands into narrow Landau levels (c). Narrow mini-bands are achieved by spatially patterning the electron system into narrow structures to yield size-quantization effects or regular arrangements thereof (d). No magnetic fields are required in this case. Electronic correlations are also generated if the material properties are altered such that, for example, local inter-electron Coulomb repulsion energies $U$ are so large as to be comparable to $W$ (e), yielding Mott–Hubbard physics.

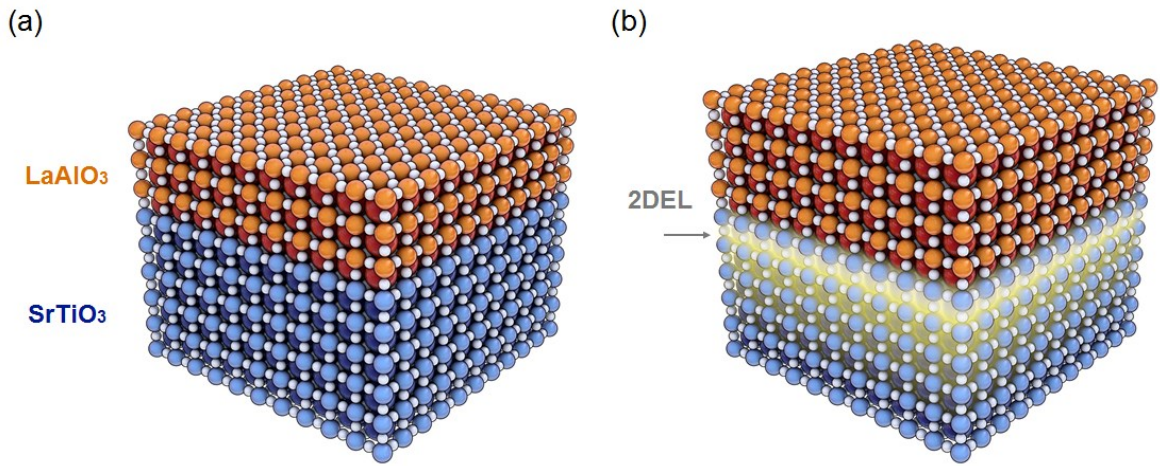

Figure 2

Heterostructures consisting of LaAlO$_3$ films grown on the TiO$_2$-terminated (001) surface of SrTiO$_3$. At the interface between these two insulators, a 2D electron liquid (yellow) is generated if the LaAlO$_3$ layer is thicker than 3 unit cells.

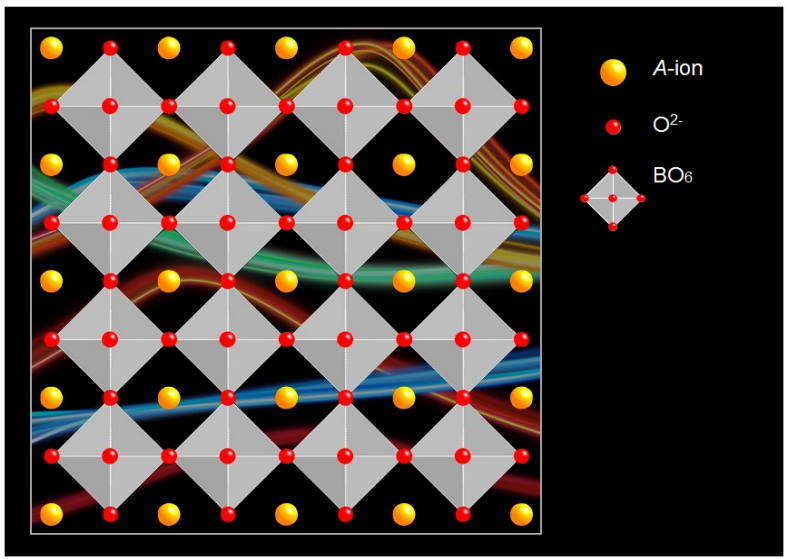

Figure 3

Illustration of an artificial atom built from a correlated material. The drawing shows the top view of a quantum dot formed by a perovskite, its BO$_6$ octahedra depicted in gray. The colored wave pattern in the background symbolizes the coherent many-electron wave function formed by the correlated system.

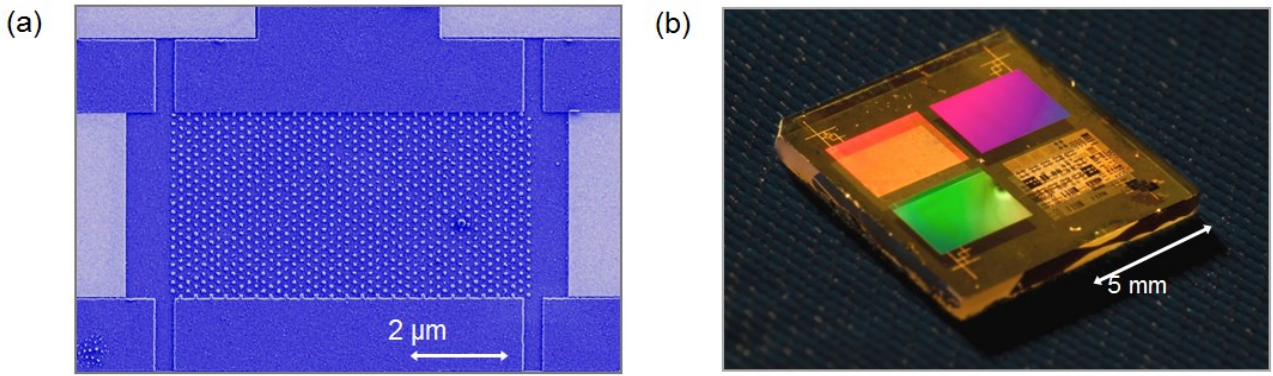

Figure 4

Images of patterned LaAlO$_3$–SrTiO$_3$ samples. Panel (a) shows an SEM micrograph of an antidot array in a 5 unit-cell-thick LaAlO$_3$ sample patterned by electron-beam lithography. The antidots have a diameter of ~90 nm and are spaced with a period of 230 nm [61]. In (b) a LaAlO$_3$–SrTiO$_3$ chip (sidelength 10 mm) is shown that contains more than 700 000 patterned field-effect transistors with feature sizes as small as 350 nm [61].

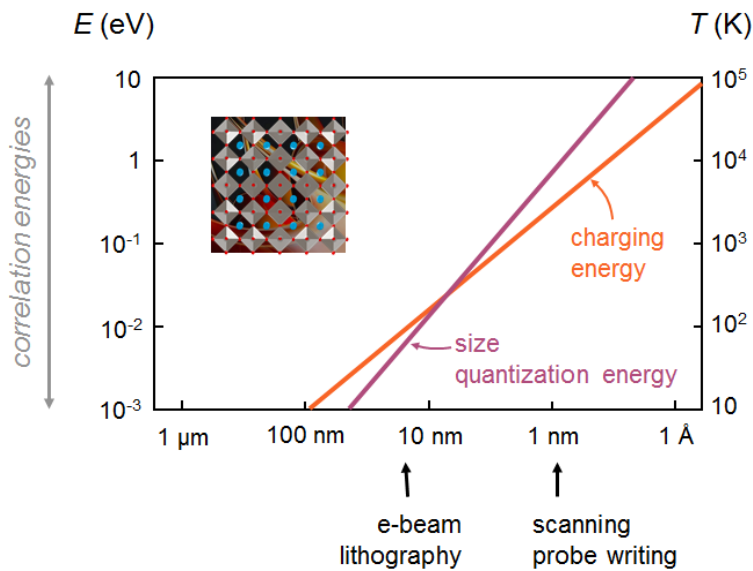

Figure 5

Dependence of the size quantization energy and the charging energy of thin quantum dots as a function of dot diameter. For these estimations, typical material parameters of complex oxides are used ($m^* = 3m_0$, $\varepsilon_r = 20$). For a large range of dot sizes, the energy scale is readily dominated by correlation energies. In contrast, the size quantization energy and the charging energy of quantum dots based on standard semiconductors are 1–2 orders of magnitude larger (not plotted) and are therefore more dominant dot properties.

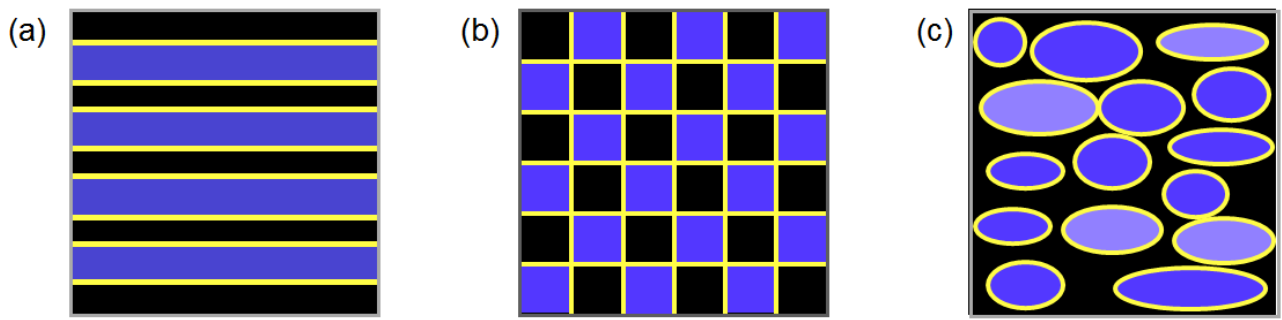

Figure 6

Illustration of bulk superconductors utilizing interfaces and quantum dots. The superconductors consist of sheets or particles of a core material (blue) embedded in a matrix material (gray), thereby forming interfaces (yellow). The matrix and core materials may be superconductors on their own. Panel (a) shows the stacking of planar interface planes, panels (b) and (c) the 2D or 3D arrangements of small particles such as quantum dots. The latter arrangements yield greater coupling in c direction than those in (a). Enhancements of the superconductivity $T_c$ are possible by means of the size effects in the quantum dots as discussed in the main text and by the interfaces (figures from [91]).

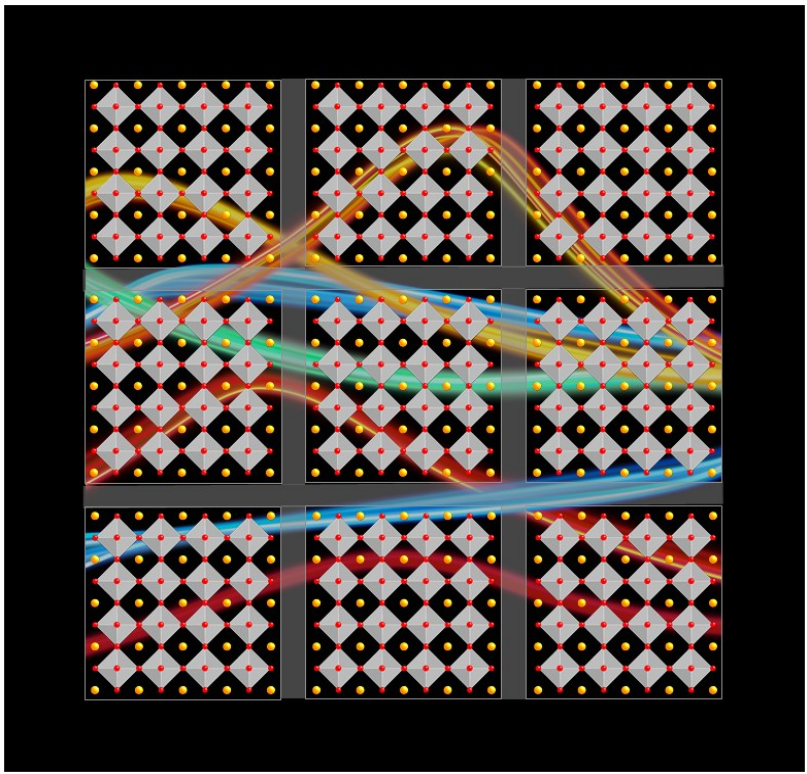

Figure 7

Illustration of a two-dimensional array of coupled artificial atoms built of quantum matter. In this sketch, the artificial atoms consist of particles of a perovskite (ABO$_3$), the BO$_6$ octahedra of which are depicted in gray (A ions orange, B ions gray, $O^{2-}$ red). Reminiscent of a superconducting state or the condensation of cold atoms into a ground state, the electron system of the array forms a coherent many-body quantum state, as symbolized by the wave pattern drawn in the background.

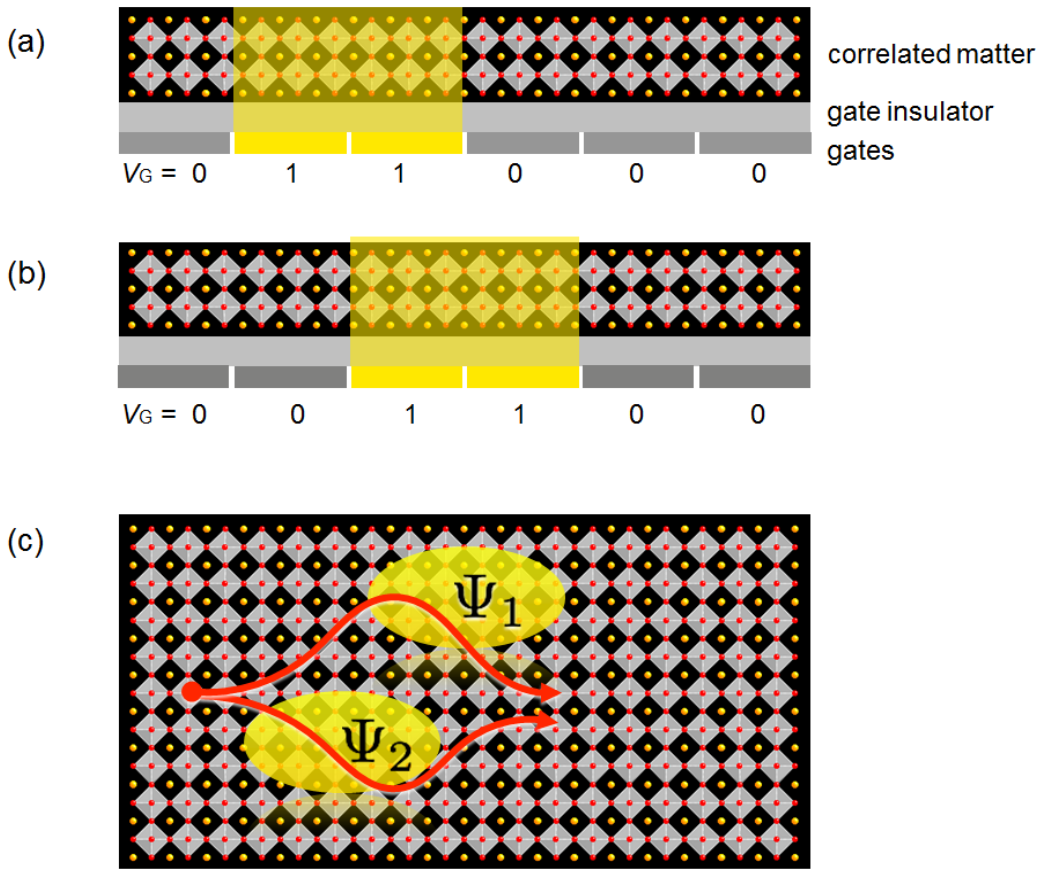

Figure 8

Illustration showing one possibility to move the many-body electron states of artificial atoms in real space. The electron states are induced in an extended, charge-depleted quantum material (a) by charging two gate electrodes (yellow). A traveling wave bias of the gates takes the electron system along with it (b). The states can also be moved in two dimensions (c) and, with restrictions, even in three.